%% file: main.tex
\documentclass[conference]{IEEEtran}

\usepackage{cite,url}
\usepackage{graphicx}
\usepackage{float}
\usepackage[caption=false]{subfig} 
\usepackage{subfig}
\usepackage{amsmath}
\usepackage{amsfonts}
\usepackage{array}
\usepackage[linesnumbered,ruled]{algorithm2e}
\usepackage[noend]{algpseudocode} 
\usepackage{gensymb}

\newcolumntype{M}[1]{>{\centering}m{#1}}
\newcolumntype{N}{@{}m{0pt}@{}}
\begin{document}
\title{Coexistence of Radar and  Communication Systems in CBRS Bands Through Downlink Power Control}
\author{\IEEEauthorblockN{
Neelakantan Nurani Krishnan\IEEEauthorrefmark{1},
Ratnesh Kumbhkar\IEEEauthorrefmark{1},
Narayan B. Mandayam\IEEEauthorrefmark{1}, 
Ivan Seskar\IEEEauthorrefmark{1} and
Sastry Kompella\IEEEauthorrefmark{2}}
\IEEEauthorblockA{\IEEEauthorrefmark{1}WINLAB, Rutgers, The State University of New Jersey, North Brunswick, NJ, USA.\\ Email: \{neel45, ratnesh, narayan, seskar\}@winlab.rutgers.edu}
\IEEEauthorblockA{\IEEEauthorrefmark{2}Information Technology Division, 
Naval Research Laboratory, Washington DC, USA\\
Email: sk@ieee.org}}

\maketitle

\begin{abstract}
Citizen Broadband Radio Service band ($3550-3700$~GHz) is seen as one of the key frequency bands to enable improvements in performance of wireless broadband and cellular systems. A careful study of interference caused by a secondary cellular communication system coexisting with an incumbent naval radar is required to establish a pragmatic protection distance, which not only protects the incumbent from harmful interference but also increases the spectrum access opportunity for the secondary system.
In this context, this paper investigates the co-channel and adjacent channel coexistence of a ship-borne naval radar and a wide-area cellular communication system and presents the analysis of interference caused by downlink transmission in the cellular system on the naval radar for different values of radar protection distance. The results of such analysis suggest that maintaining a protection distance of $30$ km from the radar will ensure the required INR protection criterion of $-6$ dB at the radar receiver with $> 0.9$ probability, even when the secondary network operates in the same channel as the radar. Novel power control algorithms to assign operating powers to the coexisting cellular devices are also proposed to further reduce the protection distance from radar while still meeting the radar INR protection requirement. 
\end{abstract}


\input{sec_intro}\vspace*{-0.2cm}
\input{sec_Analysis}
\input{sec_Interference}
\input{sec_Observations}
\vspace*{-0.2cm}
\input{sec_power_allocation}
\vspace*{-0.4cm}
\input{sec_conclusion}
\section*{ACKNOWLEDGMENT}
This work is supported in part by a grant from the U.S. Office of Naval Research (ONR) under grant number N00014-15-1-2168. The work of S. Kompella is supported directly by the Office of Naval Research.
\vspace*{-0.2cm}
\bibliographystyle{IEEEtran}
\bibliography{IEEEabrv,milcom} \end{document}

%% file: sec_intro.tex
\section{INTRODUCTION} 
\label{sec:intro}
Mobile data throughput requirement, which is highly dependent on wireless spectrum as the main resource, has been consistently rising over the last decade. It is estimated that between $2016$ and $2021$, there will be $7$x  increase in mobile data traffic \cite{ciscotraffic}. These throughput requirements have prompted  regulatory bodies such as the FCC (USA) to adopt measures like proposing policy changes to enable efficient utilization of the available spectrum, and opening up new bands or previously restricted bands for commercial unlicensed usage. To this effect, the FCC has established Citizens Broadband Radio Service (CBRS) bands for shared use of the $3550$-$3700$~MHz band ($3.5$~GHz band) for wireless broadband. Access to the CBRS bands is provided using a three-tiered spectrum authorization framework which  are: (1)~Incumbent Access (naval radar and satellites), (2)~Priority Access License (PAL), and (3)~General Authorized Access (GAA). 
The incumbent devices have the highest priority, while the PAL devices get the authorization to use the CBRS for a limited amount of time (three years), and both protected by any GAA interference. GAA devices can operate in an unlicensed manner, however no form of interference protection is provided to these devices. 
A non-incumbent device operating in the CBRS bands is commonly referred to as a Citizens Broadband Service Device (CBSD).  Shipborne naval radars are the common incumbents which operate in and adjacent to the CBRS band. In order to protect these radars, the regulatory bodies recommend protection distances which can extend even hundreds of kilometers inland from the coast \cite{locke2010assessment}.  However, this approach may severely restrict the spectrum access opportunity of the coastal areas where a large portion of US population resides \cite{paisana2014survey}. This loss of access opportunity motivates a careful analysis of this protection distance and study of  potential techniques to  minimize this loss.

In order to address this issue, this paper investigates the spectral coexistence of a shipborne naval radar and CBSDs deployed in a cell-based network, both operating in $3.5$~GHz bands. Note that, in this paper, CBSDs are equivalent to base stations in the context of a typical cellular communication network. Both in-band and adjacent band radar transmissions are considered, as specified by National Telecommunications and Information Administration (NTIA) \cite{eHata}.  The radar transmission is modeled using parameters provided by the NTIA \cite{eHata} and mathematical models provided by (International Telecommunication Union) ITU \cite{1851}. CBSD transmissions are modeled using parameters suggested by the regulatory bodies \cite{eHata}. A detailed characterization and analysis of the interference observed at the naval radar receiver due to transmissions of CBSDs is presented for varying values of radar protection distance. The interference analysis is performed in the downlink direction only as uplink transmissions (from users to CBSDs) have little contribution to the aggregate interference seen at the radar. This claim can be justified using two reasons : (a) users typically operate in low transmit powers compared to CBSDs, and/or (b) uplink from users might be in a frequency band different from the downlink, i.e, the system is FDD. To improve the spectrum access opportunity for CBSDs and to further reduce the protection distance from the radar while still meeting the performance requirement at the radar, a centralized power control mechanism to adapt the operating power of coexisting CBSDs is also presented. 
\vspace*{-0.2cm}
\subsection{Related Work}
Recently the coexistence between the  radar and cellular systems in $3.5$~GHz has been of immense interest. NTIA has made few recommendations for operation in this band regarding exclusion zone, impact of radar transmission on LTE base stations, and electromagnetic compatibility between multiple technologies \cite{eHata, ntia2014measure, ntia2014emc}. 
In \cite{sohul2015sas},  multiple spectrum access system (SAS) functionalities are presented to enable coexistence between heterogeneous systems in this band.  
Authors in \cite{ghorbanzadeh2015radar} study the impact of the transmission from shipborne S-band radar systems on the  cellular LTE systems for both co-channel and adjacent channel interference. Similar study of interference  from  the radar  is also performed in \cite{reed2016co} in an experimental setup. 
While there is an abundance of work analyzing the effect of radar on the performance of cellular communication systems, there is a lack of characterization of potential interference to the radar from the coexisting CBSDs, even if the latter adhere to the recommended regulations. This is important since such a  characterization will enable the understanding of how the interference  at the radar varies with the choice of protection distance and CBSD operating power. A similar analysis of aggregate interference  at an air traffic control radar stationed in airports due to downlink transmissions in a coexisting cell-based network is investigated in \cite{Globecom2017}. 

In this context, this paper investigates the aforementioned interference impact on radar's receiver from the downlink transmissions of coexisting cellular CBSDs. The major contribution of this paper is to provide a comprehensive characterization and analysis of the aggregate co-channel and adjacent-channel interference from downlink transmission of cellular CBSDs at the radar receiver by incorporating: i) realistic channel conditions with appropriate modeling of pathloss and large-scale fading, and ii) accurate radar beam pattern as specified by regulatory bodies, so that the modeling is more holistic and precise. This analysis is utilized to first determine an appropriate protection distance from the radar to meet a specific radar interference-to-noise ratio (INR) protection criterion, and then to propose power control algorithms to adjust the transmit power of CBSDs to facilitate further reduction of this protection distance, thereby increasing the spectrum access opportunity for the co-existing CBSDs.

%% file: sec_Analysis.tex
\section{SYSTEM MODEL} 
\label{sec:system}
\begin{figure}[t]
\centering 
\includegraphics[width=0.46\textwidth]{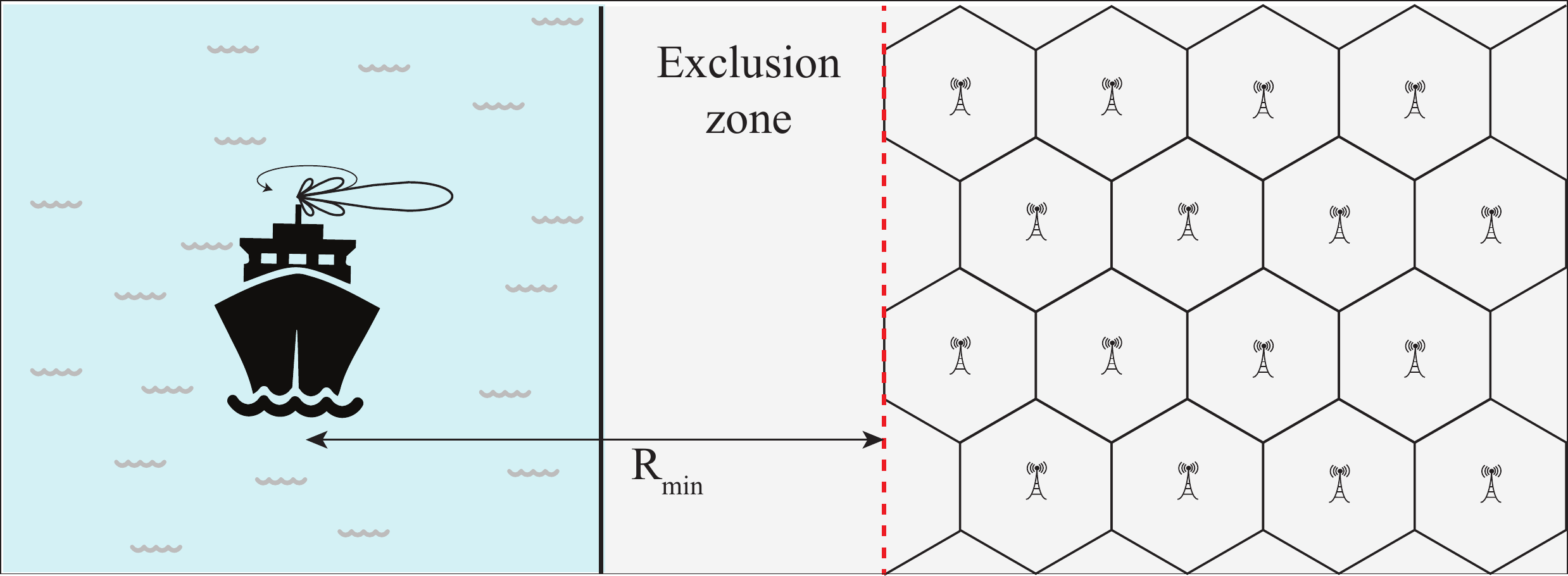}
\caption{Cellular CBSDs deployed beyond the exclusion zone in presence of shipborne naval radar; note that the interference seen at radar is contributed by only the downlink transmissions of CBSDs.}
\label{fig:scenario}
\end{figure}

This paper studies the impact of downlink transmissions from CBSDs on the performance of a naval radar. The scenario considered in this paper is shown in Fig.~\ref{fig:scenario}, where the CBSDs are deployed  beyond a  protection distance ($\text{R}_\text{min}$) from the ship.
The analysis presented in this section characterizes the impact of protection distance on the SINR performance of the radar. 
The specifications of CBSDs and channel modeling are listed in Table~\ref{specs_LTE}. The system model and associated analysis in the following sections considers both co-channel and adjacent channel operation of CBSDs and assumes that interference from CBSDs to radar is primarily in the downlink direction.
%
\begin{table}[t]
\centering
\caption{Shipborne naval radar specifications}
\begin{tabular}{|M{0.2\textwidth}|M{0.07\textwidth}|c|}
\hline 
\textbf{Operating Parameters} & \textbf{Notation} & \textbf{Value} \\ \hline \hline 
Center Frequency & $f$ & $3600$ MHz \\ 
Wavelength & $\lambda$ & $0.083$ m \\
Transmit Power & $\text{P}_\text{S}$ & $1.32$ MW \\ 
Noise Power & $\text{P}_\text{N}$ & 10$\log$(kTB) + NF dB\\
Boltzmann's constant & k & $1.38 \text{x} 10^{-23}$ J/K \\ 
Temperature & T & $290$ K \\ 
Channel Bandwidth & B & $10$ MHz \\ 
Receiver Noise Figure & NF & $3$ dB \\ 
Radar Cross Section & $\Omega$ & $100 \, \text{m}^2$ \\ 
Interference-to-Noise Ratio & $\text{INR}_\text{thr}$ & $-6$ dB \\ \hline
\textbf{Antenna Parameters} & & \\
Radiation Pattern & $\text{g}(\phi)$ & ITU-R M.1851 \\
Field Distribution & - & Cosine \\ 
Main Beam Gain & $\text{G}_\text{T}$ & $33.5$ dBi \\
3-dB Beamwidth (Azimuth) & - & $0.81^\circ$ \\ 
Side Lobe Level & - & 7.3 dBi \\
Antenna Height & $\text{h}_\text{R}$ & $8$ m \\ \hline 
\end{tabular}
\label{specifications}
\end{table} 
\begin{table}[t]
\centering 
\caption{CBSD specifications and Channel modeling}
\begin{tabular}{|M{0.21\textwidth}|c|c|}
\hline
\textbf{Operating Parameters} & \textbf{Notation} & \textbf{Value} \\ \hline \hline 
Maximum Transmit Power & $\text{P}_\text{BS}^\text{max}$ & $30$ dBm EIRP \\
Minimum Transmit Power & $\text{P}_\text{BS}^\text{min}$ & $20$ dBm EIRP \\
Antenna Height & $\text{h}_\text{BS}$ & $30$ m \\
Channel Bandwidth & B & $10$ MHz \\
Pathloss Model & $\rho$ & Extended HATA Model \\
Environment & - & Outdoor Urban \\
Fading & F & Log-normally distributed \\ 
Std deviation of fading & $\sigma$ & $8$ dB \\ \hline 
\end{tabular}
\label{specs_LTE} 
\end{table} 

%% file: sec_Interference.tex
Generally the CBSDs are located at a large distance from the naval radar (in the order of tens of kilometers) to satisfy the interference criterion. Therefore, the propagation environment and received signal strength will be predominantly impacted by pathloss and large-scale fading due to shadowing from obstacles, and not by small-scale fading. 
\vspace*{-0.2cm}
\subsection{Pathloss and Fading Model} 
\label{sec:interf:pathloss}


This work uses extended HATA (eHATA) model in point-to-point mode to compute the median transmission loss between a CBSD and the radar receiver.  For a link distance of $r$ km is the median pathloss determined by the following set of equations \cite{eHata} --- 
\begin{align*} 
\rho\text{(dB)} &= 30.52 + 16.81\log(f) + 4.45\log(f)^2  \\ &+(24.9 - 6.55\log(\text{h}_\text{BS}))\log(\text{R}_\text{bp}) + 10n\log\left(\frac{r}{\text{R}_\text{bp}}\right) \\ &+ 13.82\log\left(\frac{200}{\text{h}_\text{BS}}\right) + a(3) - a(\text{h}_\text{R}) + \text{FSL}(f,\text{R})
\end{align*}
where 
\begin{itemize} 
\item $\text{R}_\text{bp}$ - Break point distance 
\begin{equation*} 
\text{R}_\text{bp} = \left(10^{2n_h}\frac{a_\text{bm}(f,1)}{a_\text{bm}(f,100)}\right)^{\frac{1}{(n_h-n_l)}}
\end{equation*}
\item $a_\text{bm}$ - Frequency extrapolated basic median transmission attenuation with respect to free space 
\item $n_l(\text{h}_\text{BS})$ - Base station height dependence of the lower distance range power law exponent of the median attenuation relative to free space 
\begin{equation*} 
n_l(\text{h}_\text{BS}) = 0.1(24.9 - 6.55\log{\text{h}_\text{BS}})
\end{equation*}
\item $n_h$ - Higher distance range power law exponent of the median attenuation relative to free space
\item $n$ - Modified pathloss exponent 
\end{itemize}
\small
\begin{equation*} 
n = 
\begin{cases} 
0.1(24.9 - 6.55\log{\text{h}_\text{BS}}),& \text{if}\, 1\,\text{km} \leq r \leq \text{R}_\text{bp} \\ 
2(3.27\log{\text{h}_\text{BS}} - 0.67(\log{\text{h}_\text{BS}})^2 - 1.75),& \text{if}\, r \geq \text{R}_\text{bp}
\end{cases}
\end{equation*}
\normalsize
\begin{itemize}
\item a($\text{h}_\text{R}$) - Radar reference height correction 
\item FSL - Free space loss at distance R 
\begin{equation*}
\text{R} = \sqrt{(r\times10^3)^2 + (\text{h}_\text{BS} - \text{h}_\text{R})^2}
\end{equation*}
\end{itemize} 
Note that since the channel between the naval radar and CBSDs is mixed between land and water, appropriate corrections are added to the pathloss computed above as suggested by the model \cite{eHata}. 

Typical simple fading models (like exponential or Rayleigh modeling) do not accurately capture the observed fading in the link between CBDSs and radar. 
This is because the two are separated far apart from each other and hence large-scale shadowing dominates. 
This paper models the fading between the two as a log-normal random variable, which has been classically used in literature to model large-scale fading in interference analysis. 
Similarly, log-normal fading is also used to model the fading between radar and its intended target. 




\subsection{Interference Model}
\label{sec:interf:interf}
The quantity of interest at the radar receiver is the distribution of signal to interference noise ratio (SINR) as radar target detection performance is characterized by the attained SINR. SINR at radar receiver can be modeled as --- 
\begin{equation}
\label{sinr}
\text{SINR} = \frac{\frac{\text{P}_\text{S}\text{G}_\text{T}\Omega\lambda^2\text{F}_\text{R}}{(4\pi)^3\text{R}_\text{T}^4}}{\text{P}_\text{N} + \sum\limits_{m=1}^M \text{P}_\text{BS}\rho(r_\text{m})\text{F}_\text{BS}g(\phi_m)}
\end{equation}
where 
\begin{itemize}
  \setlength{\itemsep}{1pt}
  \setlength{\parskip}{0pt}
  \setlength{\parsep}{0pt}
\item $\text{R}_\text{T}$ is the distance of target from radar (radar range), 
\item $M$ denotes the number of CBSDs, 
\item $r_\text{m}$ is the distance of $\text{m}^\text{th}$ CBSD from radar receiver, 
\item $g(\phi_\text{m})$ is the gain of radar beam at the $\text{m}^\text{th}$ CBSD location, and
\item $\text{F}_\text{R}$ and $\text{F}_\text{BS}$ represent the fading random variables between radar-target and BS-radar respectively. 
\end{itemize}


The aggregate interference in the denominator of eq.~\ref{sinr} is a sum of uncorrelated log-normal random variables, which has been analytically characterized in \cite{lognormal}. The distribution of SINR at the radar receiver can hence be modeled as --- 
\begin{equation} 
\label{cdf}
\mathbb{P}[\text{SINR} < \mathcal{T}] = 1-\mathbb{Q}\left(\frac{10\log \mathcal{T}+10\log \Lambda}{\bar{\sigma}}\right)
\end{equation}
where: 
\begin{align*}
\Lambda &= \sum_{m=1}^\text{M} 10^{(t_\text{m} - t_\text{i})/10} + \frac{1}{\gamma} \\ 
t_\text{m} &= \text{P}_\text{BS} \text{(dBm)} - \rho(r_\text{m}) + g(\phi_\text{m}) \\ 
t_\text{i} &= 10\log(\text{P}_\text{R}) - 40\log(\text{R}_\text{T}) \\ 
\text{P}_\text{R} &= \frac{\text{P}_\text{S}\text{G}_\text{T}\Omega\lambda^2}{(4\pi)^3\text{R}_\text{T}^4}, \quad\quad \quad\gamma = \frac{\text{P}_\text{R}}{\text{R}_\text{T}^4\text{P}_\text{N}} \\ 
\bar{\sigma} &= \sigma^2 + \sum_{m=1}^M \lambda_k^2 \sigma^2 , \quad\lambda_m = \frac{10^{(t_\text{m} - t_\text{i})/10}}{\Lambda}
\end{align*} 
The characterization of SINR given in eq.~\ref{sinr} is when CBSD devices are operating in the same channel as the radar itself, that is, the aggregate CBSD interference is modeled as co-channel with the radar. When the CBSD devices operate in a channel adjacent to the radar transmission, the aggregate interference seen by the radar receiver will be substantially lower than the co-channel case. To model such a scenario, the denominator of eq.~\ref{sinr} needs to incorporate the frequency dependent rejection (FDR), which is defined as the amount of energy received by the radar receiver from channels adjacent to its operation. The method to compute FDR at the radar receiver in the presence of LTE (Long Term Evolution) communication operation in an adjacent band is described in \cite{eHata}. Denoting FDR as X (units in dB), the distribution of SINR at the radar receiver in the presence of adjacent channel interference can be modeled as ---  
\begin{equation}
\label{sinr_fdr}
\text{SINR} = \frac{\frac{\text{P}_\text{S}\text{G}_\text{T}\Omega\lambda^2\text{F}_\text{R}}{(4\pi)^3\text{R}_\text{T}^4}}{\text{P}_\text{N} + \sum\limits_{m=1}^M \text{P}_\text{BS}\rho(r_\text{m})10^{\text{X}/10}\text{F}_\text{BS}g(\phi_m)}
\end{equation}
\begin{figure*}[t] 
\centering 
  \subfloat[Co-channel]{\label{fig:plot_sinr:cochannel}\includegraphics[width=0.48\textwidth]{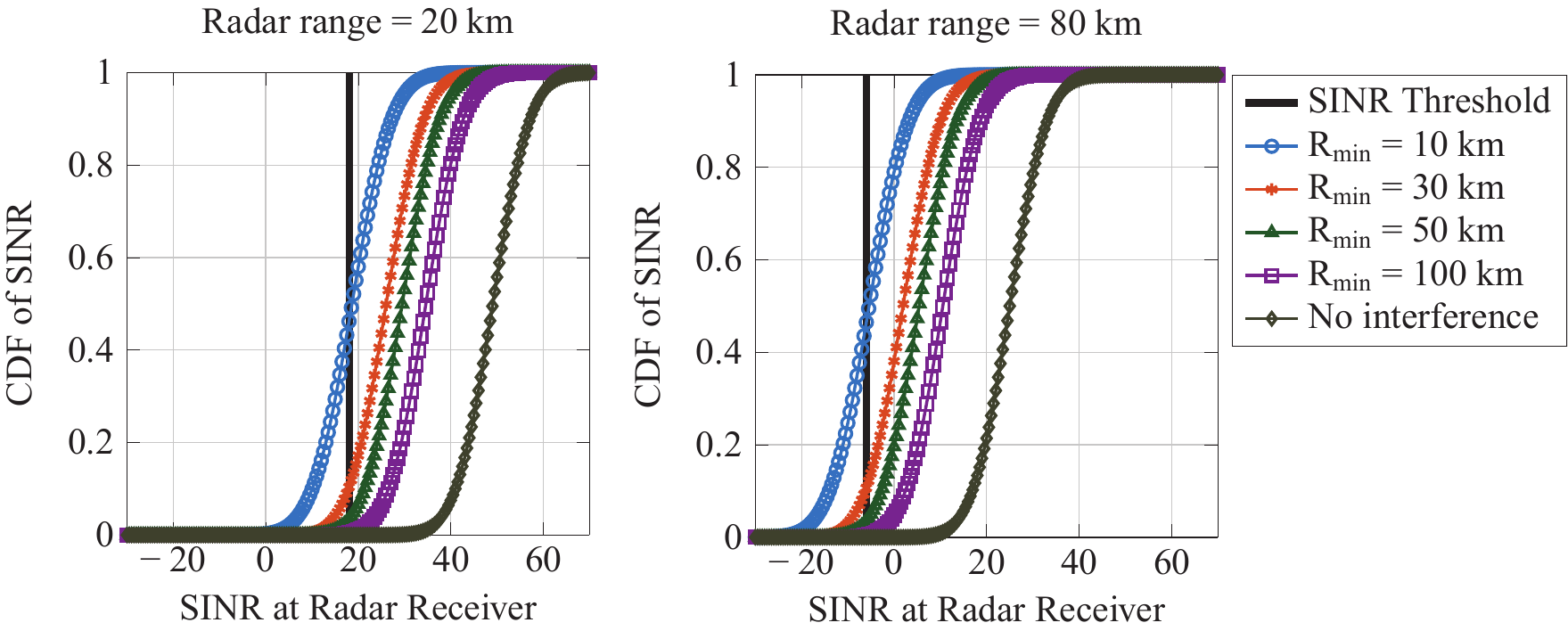}} \hfill
  \subfloat[Adjacent channel]{\label{fig:plot_sinr:adchannel}\includegraphics[width=0.48\textwidth]{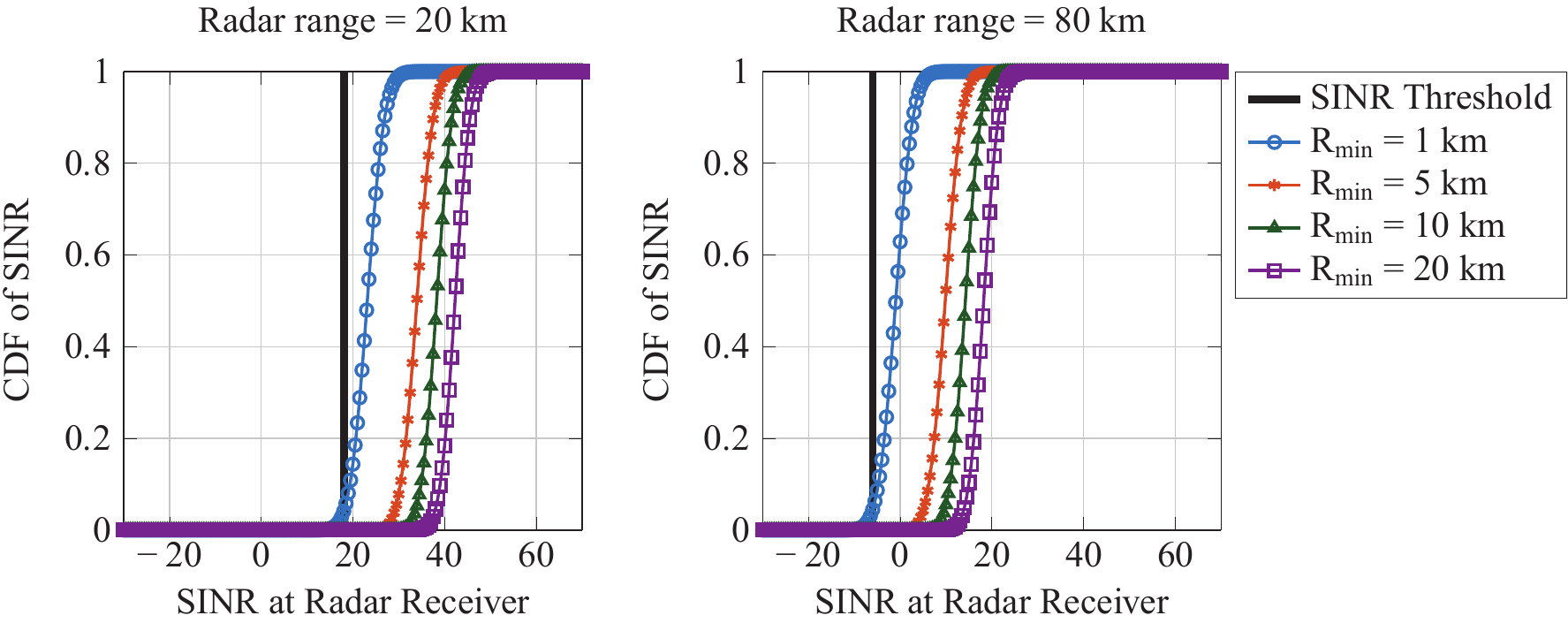}} 
  \centering
  \caption{Distribution of SINR at the radar receiver for two radar target ranges --- CBSD operation is in (a) the same channel, (b) adjacent channel as the radar}
  \label{plot_sinr}
\end{figure*}


%% file: sec_Observations.tex
\section{INTERFERENCE ANALYSIS}
\label{sec:res:main}
This section  presents the methodology used for obtaining numerical results for the analysis of aggregate interference due to CBSD transmissions (both co-channel and adjacent channel to radar) at the radar receiver and provides basic guidelines to determine the appropriate protection distance from the radar. 

\subsection{Methodology}
\label{sec:res:method}
The operating parameters of radar and CBSDs involved in the analysis are listed in Table~\ref{specifications} and \ref{specs_LTE} respectively. 
The CBSDs are deployed at distances higher than the protection distance in a cellular structure. This cellular structure is formed in such a way that the received power at one CBSD transmitting at maximum allowed power ($30$~dBm) to any other CBSD is below $-62\, \text{dBm}$. This ensure that CBSDs do not cause harmful interference to each other.
Monte Carlo simulations  over 10,000 trials are carried out by randomly deploying the CBSDs for different protection distance from radar to calculate the average distribution of SINR (eq.~\ref{cdf}) at the radar receiver. 


\subsection{Distribution of SINR - Co-channel Coexistence}
\label{sec:res:sinr}
The CDF of SINR at the radar receiver averaged over 10,000 Monte-Carlo runs for different radar target ranges and protection distances are plotted in Fig.~\ref{fig:plot_sinr:cochannel}. Note that the CBSDs operate in the same channel as the radar, that is, the interference from CBSD transmissions seen by the radar receiver is co-channel (characterization of SINR is given by eq.~\ref{sinr}). The fundamental protection criterion that needs to be achieved at the radar receiver, as specified by regulatory authorities, is an interference-to-noise ratio (INR) of at least $\text{INR}_\text{thr}=-6$ dB \cite{ntia2006inr}. In other words, radar can tolerate the interference as long as the aggregate interference is not substantial enough to violate the INR protection specification. Hence, the minimum SINR threshold to be achieved at the radar receiver corresponding to the INR requirement can be computed as 
\begin{equation}
\label{thr}
\text{SINR}_\text{thr} = \frac{\frac{\text{P}_\text{S}\text{G}_\text{T}\Omega\lambda^2}{(4\pi)^3\text{R}_\text{T}^4}}{(1 + 10^{\text{INR}_\text{thr}/10})\times\text{P}_\text{N}}
\end{equation}
where $\text{P}_\text{N}$ represents the noise power and $\text{R}_\text{T}$ denotes the radar target range. In Fig.~\ref{fig:plot_sinr:cochannel}, the SINR threshold corresponding to the corresponding radar range is plotted in bold. If the observed SINR at the radar receiver exceeds $\text{SINR}_\text{thr}$ for a certain target range and protection radius with high probability, then the interference to radar in this scenario is deemed to be unobjectionable. In the interest of completeness of analysis, distribution of SNR at the radar receiver in the absence of any CBSD interference is also plotted in Fig.~\ref{fig:plot_sinr:cochannel}. 

It can be seen consistently across the two radar ranges in Fig.~\ref{plot_sinr} that if the INR protection criterion is satisfied by a CBSD deployment at a certain protection distance, the criterion will be met for any radar target range at the same protection distance. This observation can be substantiated by the fact that aggregate interference effected by CBSD deployment on radar for a given protection distance is the same regardless of the radar target range ($\text{R}_\text{T}$), that is, the aggregate interference plus noise power is independent of $\text{R}_\text{T}$ (denominator of eq.~\ref{thr}).  
The results in Fig.~\ref{fig:plot_sinr:cochannel} indicate that if CBSD are deployed with a protection distance of $50$ km away from the radar, the effective interference at the radar receiver is within tolerable limits more than $95 \%$ of the time (i.e, with probability $0.95$). If the latter requirement (that is the percentage of time aggregate interference is within acceptable threshold) can be relaxed, the protection distance from the radar can be correspondingly shrunk to as low as $30$ km and still result in aggregate interference at the radar that is within acceptable limits more than $90\%$ of the time. Hence, such analysis and observations can provide useful guidelines regarding the choice of protection distance from the radar in order to attain a certain probability that aggregate interference must not exceed the maximum tolerable threshold. 
\vspace*{-0.5cm}
\subsection{Distribution of SINR - Adjacent channel Coexistence}
The discussions till now are under the assumption that the CBSDs operate in the same channel as the radar, that is, the radar receiver observes the aggregate interference from CBSDs in the same channel as its operation. As described in eq.~\ref{sinr_fdr}, the aggregate interference from CBSDs operating in a channel adjacent to radar operation can be characterized by accounting for the frequency dependent rejection (FDR) of the radar receiver. Using FDR values as mentioned in \cite{eHata} for a typical naval radar receiver, the distributions of SINR at the radar receiver are plotted for different radar ranges and protection distances. Comparing Fig.~\ref{fig:plot_sinr:cochannel} and Fig.~\ref{fig:plot_sinr:adchannel} for the two radar ranges, it can be observed that the protection distance is substantially lower for the adjacent channel operation than the co-channel scenario. All the major trends observed in Fig.~\ref{fig:plot_sinr:cochannel} still hold for the adjacent channel case, that is, if the INR protection criterion is satisfied by a CBSD deployment at a certain protection distance, the criterion will be met for any radar target range at the same protection distance. Fig.~\ref{fig:plot_sinr:adchannel} suggests that CBSD devices can be deployed as close as $1$ km from the radar and the resulting aggregate interference at the radar receiver will be less than the INR threshold more than $95\%$ of the time. 

%% file: sec_power_allocation.tex
\section{POWER CONTROL FOR CBSD}
\label{sec:power}

\begin{figure*}[t!] 
\centering 
  \subfloat[Protection distance = $20$ km]{\label{fig:pow_20}\includegraphics[width=0.31\textwidth]{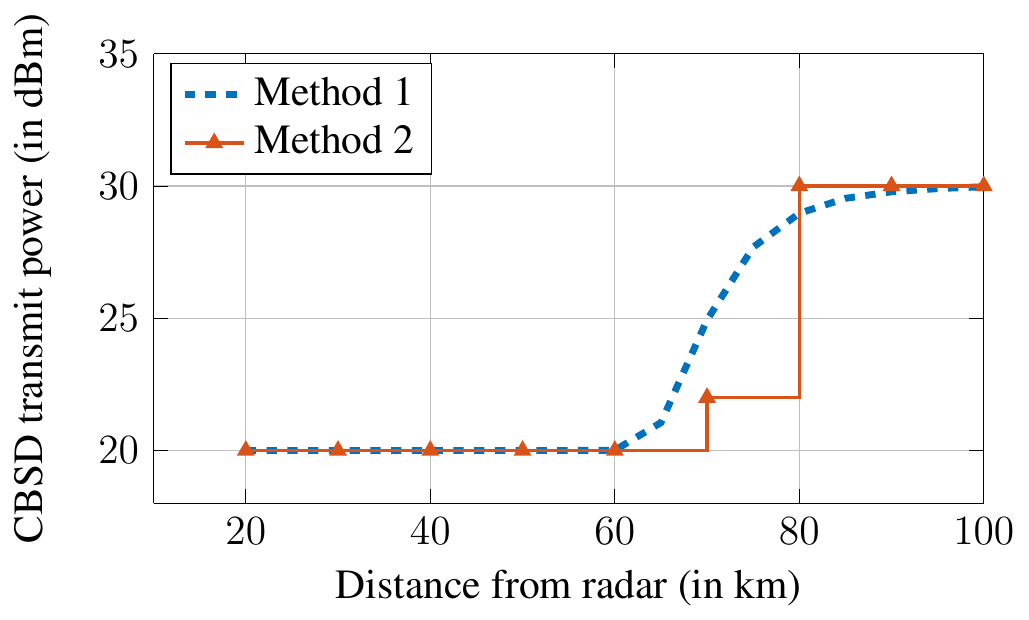}} \hfill
  \subfloat[Protection distance = $30$ km]{\label{fig:pow_30}\includegraphics[width=0.31\textwidth]{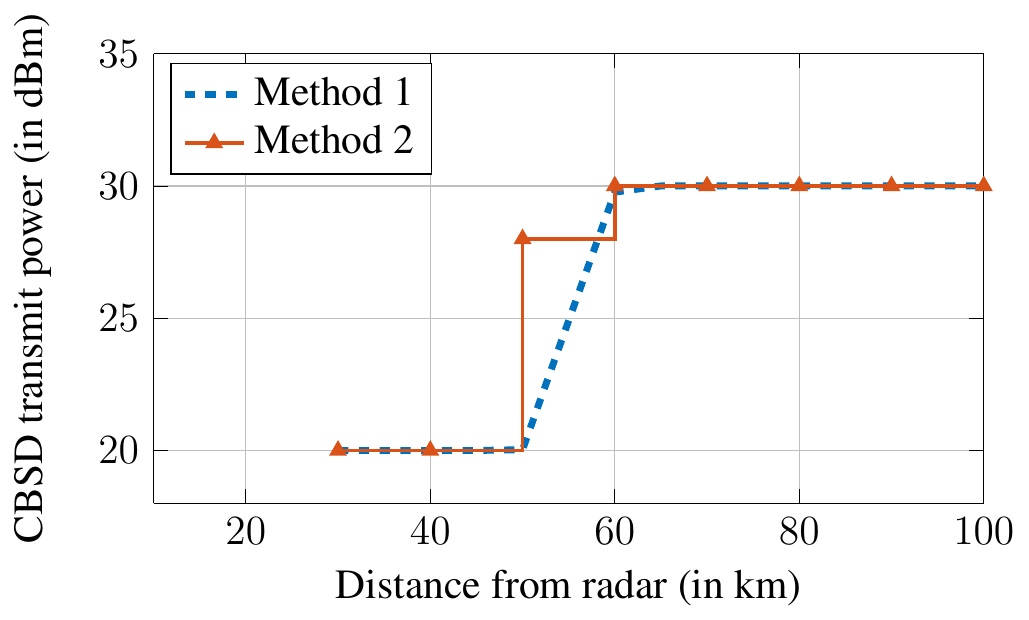}} \hfill
  \subfloat[Protection distance = $50$ km]{\label{fig:pow_50}\includegraphics[width=0.31\textwidth]{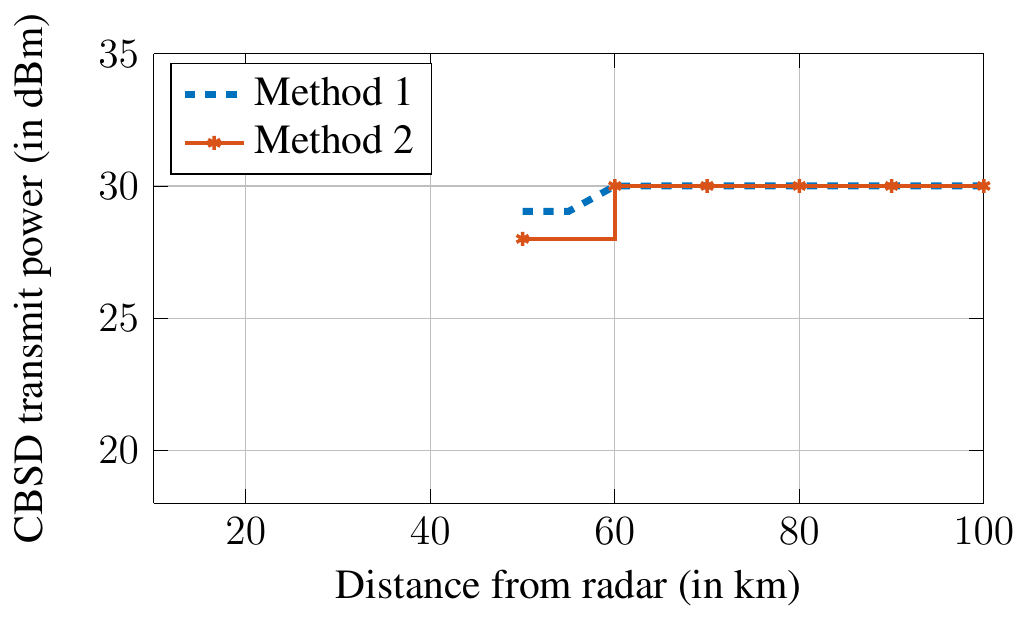}} 
  \centering
  \caption{Power distribution for CBSDs to achieve different values of protection distance; CBSDs operate in the same channel as the radar}
  \label{fig:power_co}
\end{figure*}

The analysis outlined in the previous section to determine an appropriate protection distance from the radar while meeting the INR protection criterion assumes that all the co-existing CBSDs operate at the maximum allowed transmit power of $30$~dBm. However, by enabling power control at the CBSDs to adjust their transmission power, this protection distance can potentially be further reduced and hence spectrum access opportunity can be increased for more number of CBSDs. This section presents such a power control algorithm which allocates appropriate power to the CBSDs for a given value of protection distance while still satisfying the criterion of $-6$~dB INR at the radar receiver. Note that this algorithm is centralized and requires no co-operation from the naval radar. The algorithms and corresponding results presented in subsequent sub-sections assume that the ship, and hence the radar, is located as close  to the shore as the normal operation allows, i.e, the worst-case scenario for coexistence. 

\subsection{Power control for fixed density of CBSDs}
\label{sec:power:non_dense}
\begin{figure}[t]
\centering 
\includegraphics[width=0.42\textwidth]{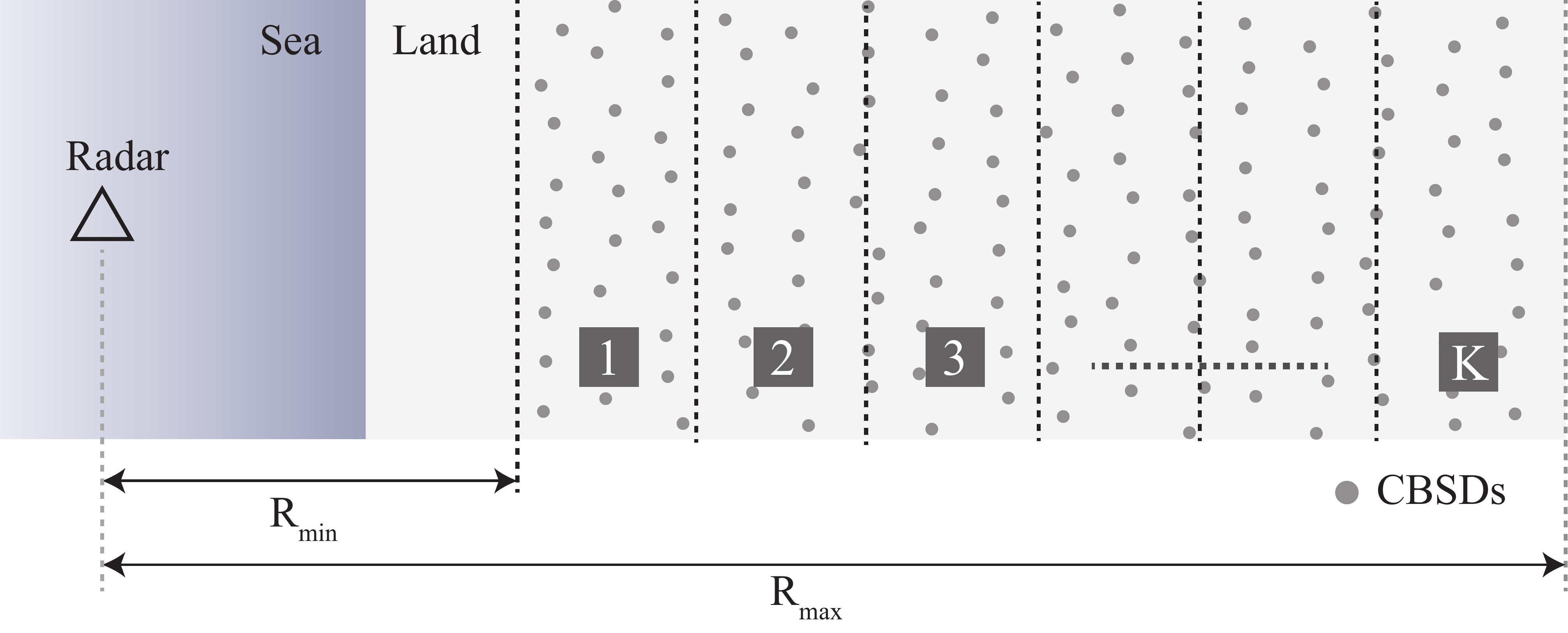}
\caption{Typical CBSD deployment scenario; the area covered by N CBSDs is divided into K sectors}
\label{fig:sectors}
\end{figure}

The power allocation scheme is presented for the scenario where the locations of CBSDs are fixed based on maximum allowed transmission power.
Refer Algorithm \ref{algo:power} for the details of the proposed power allocation scheme.  This algorithm assumes that for a particular choice of $\text{R}_\text{min}$, the CBSD transmission power can be between the limits of  $\text{P}_\text{BS}^\text{min}$ and $\text{P}_\text{BS}^\text{max}$.  Using these limits as arguments, the algorithm first defines and initializes few parameters (steps 2-5).
In the bootstrapping phase, power $\text{P}_\text{BS}^\text{min}$ is allocated to all the CBSDs and aggregate interference $\text{I}_\text{agg}$ is calculated. If, in this bootstrapping phase,  $\text{I}_\text{agg}$ exceeds the interference threshold $\text{I}_\text{th}$, then the power allocation is declared infeasible for the given value of $\text{R}_\text{min}$ and $\text{P}_\text{BS}^\text{min}$. 
In the interest of tractability of analysis, a maximum value of Radar-CBSD distance ($\text{R}_\text{max}$) is  chosen, which is a valid assumption since CBSDs beyond a certain distance would not affect the radar performance due to high pathloss suffered by transmissions from such CBSDs. 
Once the initialization phase is successfully passed, the values of operating power of the co-existing CBSDs are gradually increased, starting from CBSD located farthest from the radar in such a manner that the condition $\text{I}_\text{agg}\leq\text{I}_\text{th}$ is still met (step 9-10). This method of allocating power for individual nodes is referred to as \textit{method~1}. 
Note that, this algorithm has the complexity of $\mathcal{O}(M^2)$  for $M$ CBSDs due to iteration over all CBSDs and calculation of $\text{I}_\text{agg}$ for each iteration. In order to reduce this complexity, which is quadratic in the number of deployed CBSDs, this algorithm can be modified such that the assignment of power is performed  over a group of CBSDs instead of iterating over individual CBSD. In this context, these groups are formed by dividing the area beyond the protection distance and within the main lobe of the radar into $K\ll M$ parallel sectors as shown in Fig. \ref{fig:sectors}.  For each sector, a common value of approximate pathloss is selected for all CBSDs located within that sector and hence the complexity of the the algorithm is reduced to $\mathcal{O}(K^2)$. This method of power allocation is referred to as \textit{method 2}.

\begin{figure}
	\centering 
  	\subfloat[Fixed density]{\label{fig:power_density:non_dense}\includegraphics[height=0.12\textwidth]{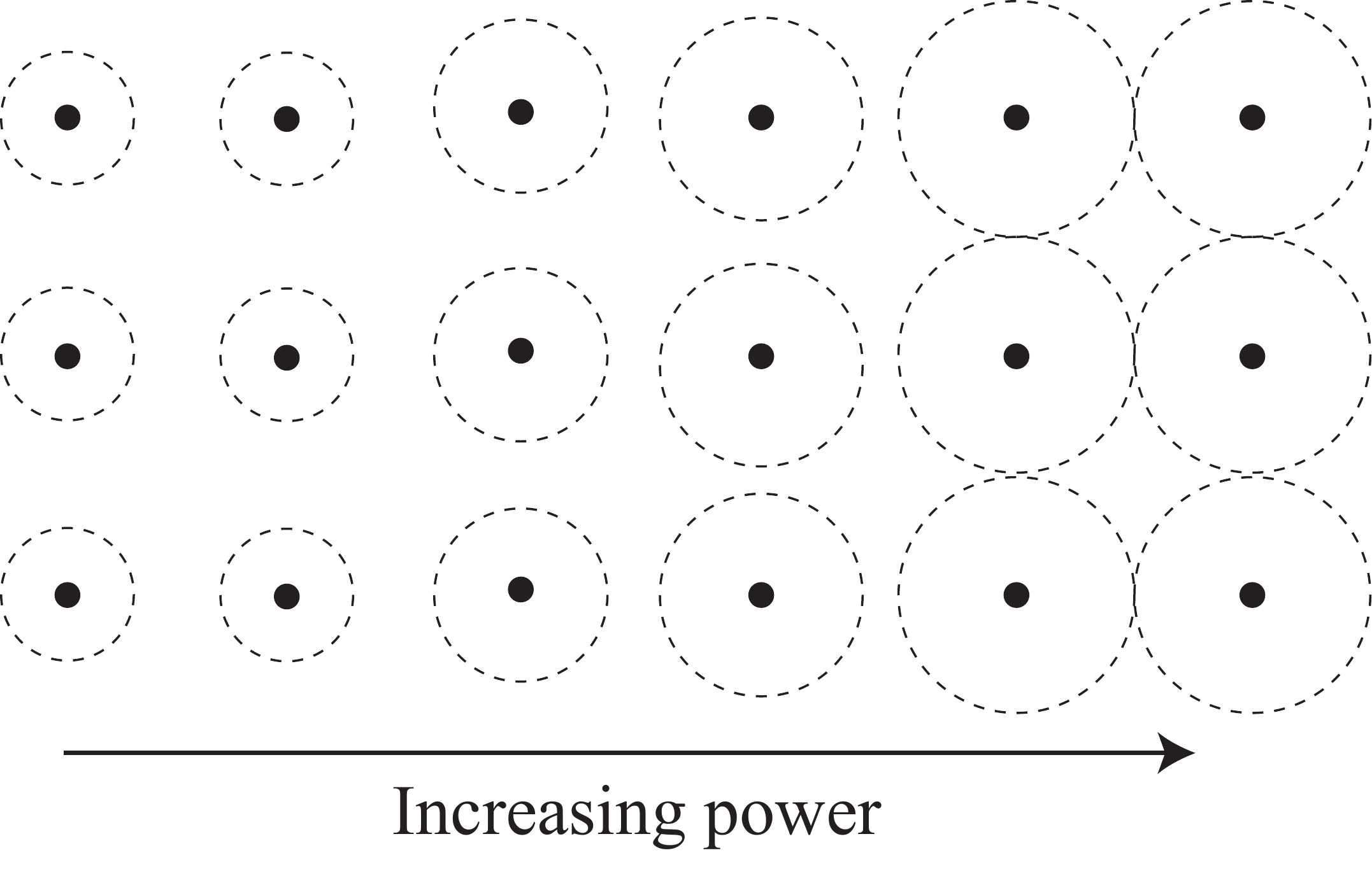}}
    \quad \quad
  	\subfloat[Density adjusted]{\label{fig:power_density:dense}\includegraphics[height=0.12\textwidth]{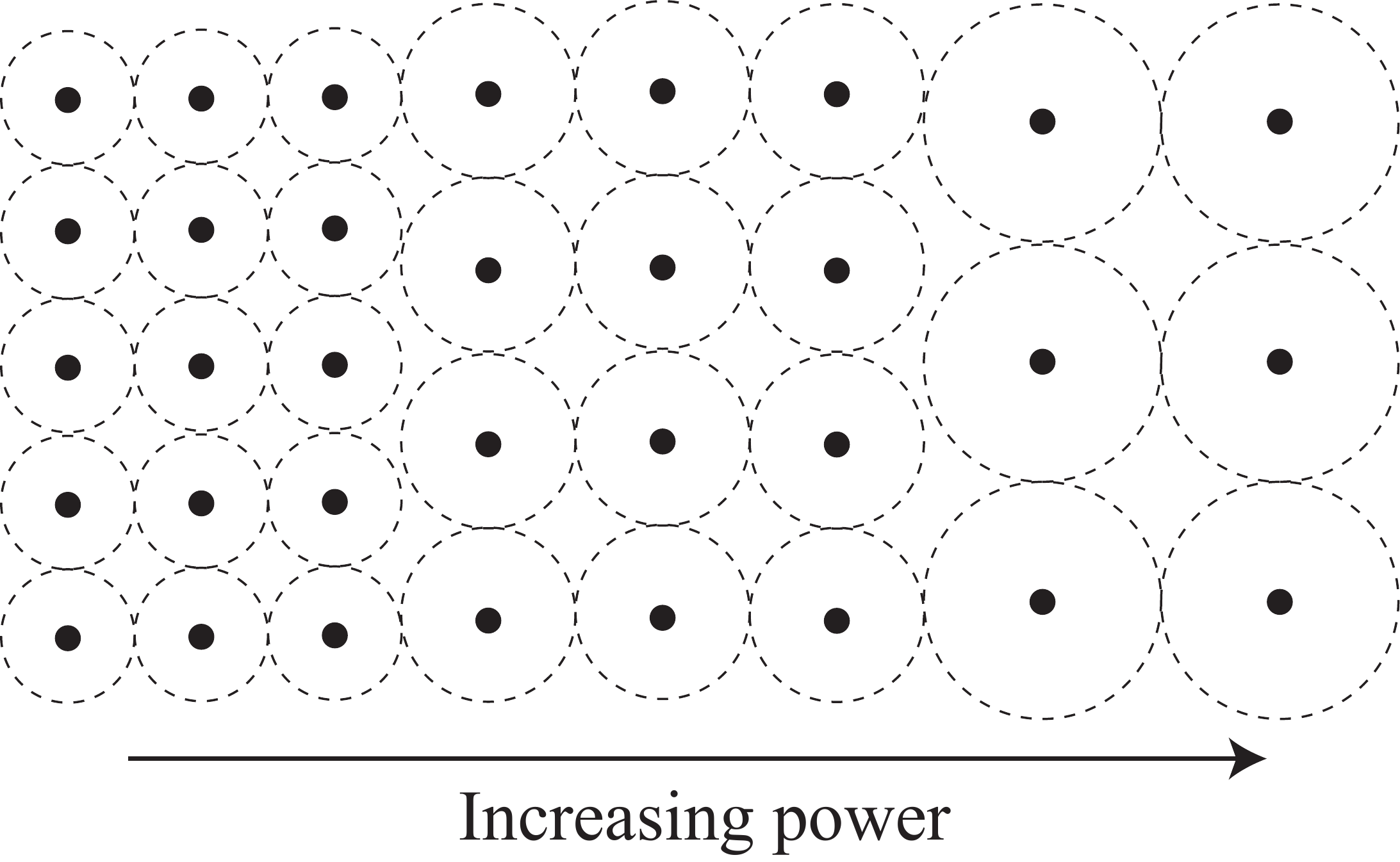}}
    \caption{Coverage area for fixed and variable densities of CBSD deployment after power control; notice coverage holes in (a)}
    \label{fig:power_density}
\end{figure}

Fig. \ref{fig:power_co} presents the results from this power allocation algorithm for  $\text{P}_\text{BS}^\text{min} = 20$~dBm,  $\text{P}_\text{BS}^\text{max}=30$~dBm, and different values of  $\text{R}_\text{min}$ . It can be observed from Fig.~\ref{fig:pow_20} that for a protection distance of $20$~km, \textit{method 1} results in power assignment of $\sim 20$~dBm for distances less than $60$~km, and $\sim 30$~dBm for the distances beyond  $90$~km. Similarly, \emph{method 2} results in a comparable power allocation of $\sim 20$~dBm for distances of less than $70$~km, and of $\sim 30$~dBm for distances beyond  $80$~km. Fig. \ref{fig:pow_30} and \ref{fig:pow_50} present corresponding results for $\text{R}_\text{min} = 30$~km and $\text{R}_\text{min}=50$~km. 
As expected, for smaller protection distances, more number of CBSDs are allocated albeit operating in lower transmit power to reduce the aggregate interference at the radar. These results substantiate the SINR analysis from previous section, in which a protection distance of $\sim 50$~km was deemed to be unobjectionable when all the CBSDs were operating at maximum allowed power. It should be noted that, since the location of CBSDs are fixed, change in transmission power results in a change in coverage range of the CBSDs as well. As an illustration, consider a typical deployment scenario depicted pictorially in Fig. \ref{fig:power_density:non_dense}, where dotted circles represent the coverage attained by a CBSD. It can be observed that due to the fixed locations of CBSD deployment and difference in transmit powers, coverage holes are formed in the network, i.e, there are pockets of regions where there is no CBSD coverage. 

\input{sec_algo}

\vspace*{-0.4cm}
\subsection{Power control with CBSD density adjustment}
\label{sec:power:non_dense}
\begin{figure}
	\centering 
  	\subfloat[Protection distance = $30$~km]{\label{fig:dense_power_density:30}\includegraphics[width=0.22\textwidth]{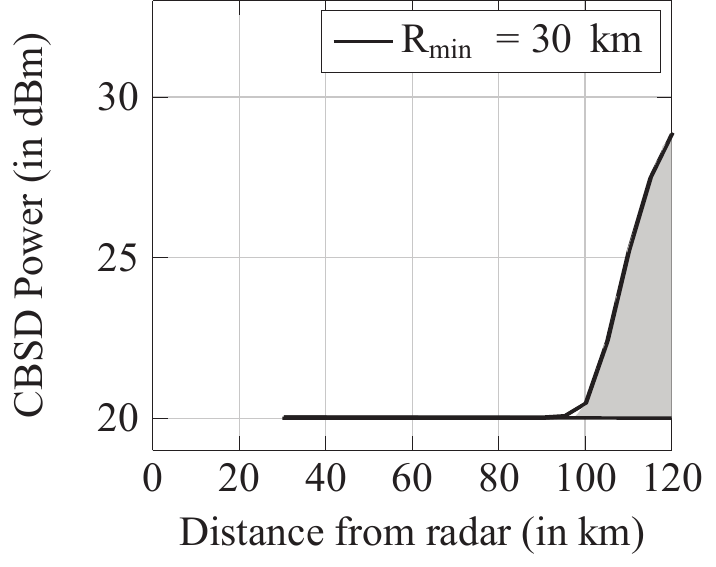}}
    \quad \quad
  	\subfloat[Protection distance = $40$~km]{\label{fig:dense_power_density:40}\includegraphics[width=0.22\textwidth]{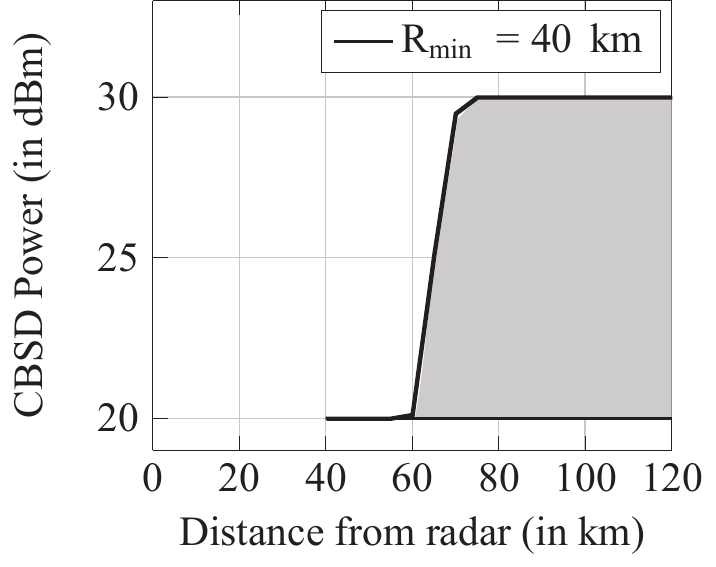}}
    \caption{Power distribution for CBSDs with density adjustment. Area highlighted in gray represents feasible values for allocated power.} \label{fig:dense_power_density}
\end{figure}

This subsection addresses the loss of coverage due to varying values of transmit power by correspondingly adjusting the device deployment density. To avoid coverage holes, the aforementioned algorithm is modified such that whenever the power allocation for CBSDs is changed in any sector, the CBSD deployment density in that sector is also correspondingly updated. Fig.\ref{fig:power_density:dense} represents this  scenario where the cell size is modified along with the transmission power. 
Similar to Algorithm 1, the modified algorithm initializes all sectors with the minimum allowed transmission power and correspondingly high CBSD density for a desired protection distance. After the initialization phase, the values of transmission power of  CBSDs are gradually increased in steps and density is correspondingly adjusted starting from the sector located farthest from the radar, while still satisfying the interference condition $\text{I}_\text{agg}\leq\text{I}_\text{th}$.
Fig. \ref{fig:dense_power_density} presents the results with this power allocation algorithm, which suggest that CBSDs can be allocated any transmission power within the highlighted region so that maximal geographic coverage is attained while still meeting the INR requirement at the radar. It can be observed from Fig. \ref{fig:dense_power_density:30} that for a protection distance of $30$~km, all CBSDs between $30$~km and $87$~km can operate with only $20$~dBm power, and CBSDs located beyond  $87$~km can be allocated any transmit power within the highlighted region as long as the corresponding CBSD density is adjusted. Similarly, it can be seen from Fig. \ref{fig:dense_power_density:30} that for a protection distance of $40$~km, CBSDs in the range of $40$~km and $60$~km can operate with  $20$~dBm power, and the CBSDs located beyond  $60$~km can be allocated  any transmit power within the highlighted region.
These results indicate that it is possible for CBSDs  to operate in small-cell or femto-cell environment without any loss of coverage while keeping the interference to the radar below the desired threshold. 

%% file: sec_algo.tex
\begin{algorithm}
    \SetKwInOut{Input}{Input}
    \SetKwInOut{Output}{Output}

    \underline{function controlBSPower} ($\text{P}_\text{BS}^\text{min}$,$\text{P}_\text{BS}^\text{max}$)\;
    $\text{P}_\text{BS} \gets $ CBSD transmit power \\
    $\text{I}_\text{agg} \gets $ Aggregate interference effected on RADAR by CBSD transmission\\
    $\text{I}_\text{th} \gets $ Maximum permitted interference (based on $\text{INR}_\text{min}$) = $-117$ dBm \\
    $\text{R}_\text{min} \gets$ Minimum protection radius around radar \\
    \textbf{Bootstrapping} - Assign $\text{P}_\text{BS}^\text{min}$ to all CBSDs \\
    Compute $\text{I}_\text{agg} = 10\log(\sum\limits_{m=1}^M 10^{\text{I}_m/10})$ where $\text{I}_m = \text{P}_\text{BS} (\text{dBm}) + \text{Radar beam gain} - \text{Path loss} - \text{Fading loss}$ \\ 
    \While{$\text{I}_\text{agg} \leq \text{I}_\text{th}$} 
      { 
        Start increasing $\text{P}_\text{BS}$ (such that $\text{P}_\text{BS} \leq \text{P}_\text{BS}^\text{max}$) of the CBSDs from the farthest end away from the radar \\
         Recompute $\text{I}_\text{agg}$ from step 6 \\  
      }
    \caption{Algorithm to adaptively control CBSD transmit power}
    \label{algo:power}
\end{algorithm}

%% file: sec_conclusion.tex
\section{CONCLUSION} 
\label{sec:conclude} 
This paper studies the coexistence of a ship-borne (naval) radar with a wide area wireless communication network composed of CBRS devices (CBSDs). A comprehensive characterization and analysis of SINR at the radar receiver is presented in presence of CBSD downlink transmissions for two cases - CBSDs operating in the same channel as the radar (co-channel interference), and in a channel adjacent to radar  (adjacent-channel interference). Monte Carlo analysis of the derived characterization reveals that CBSDs can operate as close as $30$ km away from the naval radar and still meet the INR requirement of -6 dB at the radar more than $90\%$ of the time, even when the CBSDs are operating in the same channel as the radar and transmitting at the maximum allowed power ($30$ dBm). This distance is reduced to as much as $1$ km when the CBSDs operate in a channel adjacent to the radar transmission. To further decrease the protection distance in case of co-channel CBSD operation, a novel power control algorithm is proposed which allows the CBSDs to come closer to the radar by decreasing their transmission power to meet the INR requirement at the radar. The algorithm is further modified to compensate the loss of coverage when CBSDs operate at lower transmission powers by adapting the density of deployment of CBSDs corresponding to their transmit power. 
The analysis and power control algorithms presented in this paper can be used as a general framework to choose a protection distance from any kind of radar and/or transmit power of coexisting CBSDs to meet the specified INR requirement at the radar receiver, as long as the radar specifications are known. Future work and extensions of this work may include investigating MAC-level scheduling algorithms to facilitate harmonious coexistence between these two systems, and the impact of naval radar transmission on the throughput performance of CBSDs. 